\documentclass[twocolumn,aps,prd,groupedaddress,nofootinbib,showpacs]{revtex4}
\usepackage{graphicx}
\usepackage{amsmath}
\usepackage{amssymb}
\usepackage{multirow}
\usepackage{bm}
\usepackage{color}

\usepackage{relsize}
\RequirePackage{xspace}
\usepackage{dcolumn}
\usepackage{bm}

\def\jpsi{{J/\psi}}
\def\psip{{\psi^\prime}}
\def\chicj{{\chi_{cJ}}}
\def\chicz{{\chi_{c0}}}
\def\chico{{\chi_{c1}}}
\def\chict{{\chi_{c2}}}

\def\so{{\bigl.^3\hspace{-1mm}S^{[8]}_1}}
\def\pjs{{\bigl.^3\hspace{-1mm}P^{[1]}_J}}

\def\p0{{\bigl.^3\hspace{-1mm}P^{[8]}_0}}
\def\p01{{\bigl.^3\hspace{-1mm}P^{[1]}_0}}

\def\as{\alpha_S}

\def\be{\begin{equation}}
\def\ee{\end{equation}}
\def\bea{\begin{eqnarray}}
\def\eea{\end{eqnarray}}

\def\a{\alpha}
\def\th{\theta}

\begin{document}


\title{\mbox{}\\[10pt]
Polarizations of $\chico$ and $\chict$ in Prompt Production at the
LHC}
\author{Hua-Sheng Shao$^{(a)}$, Yan-Qing Ma$^{(b)}$, Kai Wang$^{(a)}$, Kuang-Ta Chao$^{(a,c,d)}$}
\affiliation{ {\footnotesize (a)~School of Physics and State Key
Laboratory of Nuclear Physics and Technology, Peking University,
 Beijing 100871, China}\\
{\footnotesize (b)~Physics Department, Brookhaven National Laboratory, Upton, NY 11973, USA}\\
{\footnotesize (c)Collaborative Innovation Center of Quantum Matter, Beijing 100871, China}\\
{\footnotesize (d)~Center for High Energy Physics, Peking
University, Beijing 100871, China}}
\begin{abstract}
Prompt $\chi_c$ production at hadron colliders may provide a unique
test for the color-octet mechanism in nonrelativistic QCD. We
present an analysis for the polarization observables of $\chico$ and
$\chict$ at next-to-leading order in $\a_S$, and propose to measure them at the LHC,
which is expected to be important for testing the validity of NRQCD.
\end{abstract}
\maketitle
Heavy quarkonium production provides an ideal labortary to understand
quantum chromodynamics. In contrast to the helicity-summed cross
section, the quarkonium polarization measurement may
provide more complete information for the production mechanism of
heavy quarkonium~\cite{Brambilla:2010cs}.

A distinct example is the $\jpsi$ polarization at hadron
colliders. The polar asymmetry coefficient $\lambda_{\th}$ in the
angular distribution of the leptons from the $\jpsi$ decay is an
important observable that encodes the $\jpsi$ polarization
information. At the Tevatron, the CDF Collaboration measured the
quantity many years ago~\cite{Affolder:2000nn,Abulencia:2007us}.
Their measurements show that $\lambda_{\th}$ for prompt $\jpsi$
production in its helicity frame is around zero up to $p_T=30\rm{GeV}$,
indicating that the $\jpsi$ mesons are produced in the unpolarized pattern. 
The state-of-the-art theory that describes the heavy
quarkonium dynamics, non-relativistic QCD
(NRQCD)~\cite{Bodwin:1994jh}, predicts that the heavy quark pair is
allowed to be created in a color-octet (CO) intermediate state at
short distances and then evolves nonpertubatively into a
color-singlet (CS) quarkonium at long distances. Although this CO
mechanism provides an opportunity to account for the CDF yield
data~\cite{Abe:1997jz,Abe:1997yz} that cannot be resolved in the CS
model (CSM) even by including the higher-order QCD
corrections~\cite{Campbell:2007ws,Lansberg:2008gk}, the leading-order
(LO) in $\as$ NRQCD prediction gives a completely transverse
polarization result at high $p_T$ due to gluon fragmentation
to the CO $\so$ intermediate
state~\cite{Braaten:1999qk}. Recently, three groups reported
their next-to-leading order (NLO) QCD corrections to the
$\jpsi$
polarization~\cite{Butenschoen:2012px,Chao:2012iv,Gong:2012ug}.
Recall that the $\jpsi$ polarization is strongly dependent on the
specific choice of the nonperturbative long-distance matrix
elements (LDMEs), which can only be determined from the experimental
data. Choosing different $p_T$ regions of the input experimental data  may result in very different predictions.  Therefore, the
precise measurement of polarization, especially at high $p_T$, may provide a
smoking-gun signature to distinguish between various production mechanisms
of heavy quarkonium. Moreover, it was pointed out in
Ref.\cite{Chao:2012iv} that there is still a CO LDMEs parameter
space left to make both the helicity-summed yields and
$\lambda_{\th}$ quite satisfactory compared to the
hadroproduction data.

However, the prompt $\jpsi$ production at the Tevatron and LHC is affected substantially by the higher charmonia (e.g.
$\chi_c$ and $\psip$) transitions to $\jpsi$. Furthermore, even for
direct $\jpsi$ production there are three leading CO LDMEs, which
makes the precise determination of CO LDMEs
difficult.
In contrast, for the $\chi_c$ hadroproduction the feed-down
contribution only comes from $\psip$ to $\chi_c$ transition, but they are not significant,
and there is only one leading CO state $\so$
involving $\chi_c$ direct production, which can make the
determination of the LDMEs easier and
more precise. Moreover, the higher-order QCD corrections to the conventional P-wave CS
state suffer from severe infrared divergences, while in NRQCD these divergences can be absorbed by the CO state
and, thus, make the P-wave observables well defined beyond LO.
Given these reasons, the
investigation of  $\chi_c$ production at the LHC is an important
way to test the validity of NRQCD factorization and the CO mechanism.

The first investigation for the helicity-summed $\chi_c$
hadroproduction at NLO level was performed in Ref.\cite{Ma:2010vd}.
In this Letter, we extend our calculation to the polarized case, with
the method described in Refs.\cite{Chao:2012iv,Shao:2012iz}. The
polarization observables of $\chico$ and $\chict$ were
proposed in Refs.\cite{Kniehl:2003pc,Faccioli:2011be,Shao:2012fs}.
Experimentally, one may have two ways to measure the polarization
of $\chico$ and $\chict$ through the angular distributions of
their decay products. One is to measure the $\jpsi$ angular
distribution from $\chi_c\to\jpsi\gamma$. The angular distribution
with respect to the $\jpsi$ polar angle $\th$ in the rest frame of
$\chi_c$ can be formulated as~\cite{Shao:2012fs}
\begin{eqnarray}
\frac{\rm{d}\mathcal{N}^{\chicj}}{\rm{d}\cos{\th}}&\propto&
1+\sum^{J}_{k=1}{\lambda_{k\th}\cos^{2k}{\th}},
\end{eqnarray}
where the polar asymmetry coefficients $\lambda_{k\th}$ can be
expressed as the rational functions of the $\chicj$ production
spin density matrix $\rho^{\chicj}$. More specifically, for
$\chico$ it is
\begin{eqnarray}
\lambda_{\th}&=&(1-3\delta)\frac{N_{\chico}-3\rho^{\chico}_{0,0}}{(1+\delta)N_{\chico}+(1-3\delta)\rho^{\chico}_{0,0}},\label{eq:th1}
\end{eqnarray}
with $N_{\chico}\equiv
\rho^{\chico}_{1,1}+\rho^{\chico}_{0,0}+\rho^{\chico}_{-1,-1}$,
whereas for $\chict$, the coefficients are
\begin{eqnarray}
\lambda_{\th}&=&6[(1-3\delta_0-\delta_1)N_{\chict}\nonumber\\&-&(1-7\delta_0+\delta_1)(\rho^{\chict}_{1,1}+\rho^{\chict}_{-1,-1})\nonumber\\&-&(3-\delta_0-7\delta_1)\rho^{\chict}_{0,0}]/R,\nonumber\\
\lambda_{2\th}&=&(1+5\delta_0-5\delta_1)[N_{\chict}-5(\rho^{\chict}_{1,1}+\rho^{\chict}_{-1,-1})\nonumber\\&+&5\rho^{\chict}_{0,0}]/R,\label{eq:th2}
\end{eqnarray}
with
\begin{eqnarray*}
N_{\chict}&=&\rho^{\chict}_{2,2}+\rho^{\chict}_{1,1}+\rho^{\chict}_{0,0}+\rho^{\chict}_{-1,-1}+\rho^{\chict}_{-2,-2},\nonumber\\
R&=&(1+5\delta_0+3\delta_1)N_{\chict}\nonumber\\
&+&3(1-3\delta_0-\delta_1)(\rho^{\chict}_{1,1}+\rho^{\chict}_{-1,-1})\nonumber\\
&+&(5-7\delta_0-9\delta_1)\rho^{\chict}_{0,0}.
\end{eqnarray*}
The parameters $\delta$, $\delta_0$ and $\delta_1$ can be determined
by the normalized multipole amplitudes. Following the notations in
Ref.\cite{Artuso:2009aa}, we denote the normalized electric dipole
(E1) transition amplitudes by $a^{J=1}_1$ and $a^{J=2}_1$ for
$\chico$ and $\chict$, respectively, while
$a^{J=1}_2, a^{J=2}_2, a^{J=2}_3$ are the $\chico$ and $\chict$
normalized magnetic quadrupole (M2) amplitudes and $\chict$
electric octupole amplitude (E3). We remind readers that the word ``normalized" here means we have
relations $a^{J=1}_1+a^{J=1}_2=1$ and $a^{J=2}_1+a^{J=2}_2+a^{J=2}_3=1$. The explicit expressions for
$\delta, \delta_0, \delta_1$ are
\begin{eqnarray}
\delta&=&(1+2a^{J=1}_1a^{J=1}_{2})/2,\nonumber\\
\delta_0&=&[1+2a^{J=2}_1(\sqrt{5}a^{J=2}_2+2a^{J=2}_3)\nonumber\\&+&4a^{J=2}_2(a^{J=2}_2+\sqrt{5}a^{J=2}_3)+3(a^{J=2}_3)^2]/10,\nonumber\\
\delta_1&=&[9+6a^{J=2}_1(\sqrt{5}a^{J=2}_2-4a^{J=2}_3)\nonumber\\&-&4a^{J=2}_2(a^{J=2}_2+2\sqrt{5}a^{J=2}_3)+7(a^{J=2}_3)^2]/30.
\end{eqnarray}
An alternative way to study the polarizations of
$\chico$ and $\chict$ is to measure the dilepton angular
distributions from $\chicj\to\jpsi\gamma\to l^+l^-\gamma$. There are
two choices to describe the dilepton angular
distributions~\cite{Faccioli:2011be,Shao:2012fs}. Here, we only choose the second
one presented in Ref.\cite{Shao:2012fs}, where the $z$ axis in
the rest frame of $\jpsi$  coincides with the direction of
the spin quantization axis in the $\chi_c$ rest frame. The generic
lepton polar angle $\th^{\prime}$ dependence is
\begin{eqnarray}
\frac{\rm{d}\mathcal{N}^{\chicj}}{\rm{d}\cos{\th^{\prime}}}&\propto&
1+\lambda_{\th^{\prime}}\cos^{2}{\th^{\prime}},
\end{eqnarray}
where
\begin{eqnarray}
\lambda^{\chico}_{\th^{\prime}}&=&\frac{-N_{\chico}+3\rho^{\chico}_{0,0}}{R_1},\nonumber\\
\lambda^{\chict}_{\th^{\prime}}&=&\frac{6N_{\chict}-9(\rho^{\chict}_{1,1}+\rho^{\chict}_{-1,-1})-12\rho^{\chict}_{0,0}}{R_2},\label{eq:thp}
\end{eqnarray}
with
\begin{eqnarray*}
R_1&=&[(15-2(a^{J=1}_2)^2)N_{\chico}\nonumber\\&-&(5-6(a^{J=1}_2)^2)\rho^{\chico}_{0,0}]/(5-6(a^{J=1}_2)^2),\nonumber\\
R_2&=&[2(21+14(a^{J=2}_2)^2+5(a^{J=2}_3)^2)N_{\chict}\nonumber\\&+&3(7-14(a^{J=2}_2)^2-5(a^{J=2}_3)^2)(\rho^{\chict}_{1,1}+\rho^{\chict}_{-1,-1})\nonumber\\
&+&4(7-14(a^{J=2}_2)^2-5(a^{J=2}_3)^2)\rho^{\chict}_{0,0}]\nonumber\\&\div&[7-14(a^{J=2}_2)^2-5(a^{J=2}_3)^2].
\end{eqnarray*}
Note that
$\lambda_{2\th}$ for $\chict$ is suppressed by the higher-order
multipole amplitudes $a^{J=2}_2,a^{J=2}_3$. The observable is
expected to be near zero. Hence, we refrain from establishing the $p_T$
distribution of $\lambda_{2\th}$ here.

In our numerical computation, we choose the same input parameters as
those presented in Ref.\cite{Chao:2012iv}. The renormalization
scale $\mu_r$, factorization scales $\mu_f$ and NRQCD scale
$\mu_{\Lambda}$ are chosen as $\mu_r=\mu_f=\sqrt{4m_c^2+p_T^2}$ and
$\mu_{\Lambda}=m_c$. The CO LDMEs are chosen to be
$\langle\mathcal{O}^{\chicj}(\so)\rangle=(2J+1)\times(2.2^{+0.48}_{-0.32})\times10^{-3}\rm{GeV}^3$~\cite{Ma:2010vd},
which are obtained by fitting the ratio
$\sigma_{\chict}/\sigma_{\chico}$ at NLO level to the CDF
data~\cite{Abulencia:2007bra}, while the CS LDMEs are estimated using
the B-T potential model~\cite{Eichten:1995ch} as
$\langle\mathcal{O}(\pjs)\rangle=(2J+1)[(3|R^{\prime}(0)|^2)/4\pi]$
with $|R^{\prime}(0)|^2=0.075\rm{GeV}^5$. The uncertainties due to the
scale dependence, which is estimated by varying $\mu_r,\mu_f$ by a
factor of $\frac{1}{2}$ to $2$ with respect to their central values,
the charm quark mass $m_c=1.5\pm0.1\rm{GeV}$ and the error in the
CDF data~\cite{Abulencia:2007bra} are all encoded in the error
estimations of the CO LDMEs. The normalized multipole amplitudes
used here are taken from the CLEO
measurement~\cite{Artuso:2009aa}, i.e.
$a^{J=1}_2=(-6.26\pm0.68)\times10^{-2},a^{J=2}_2=(-9.3\pm1.6)\times10^{-2},a^{J=2}_3=0$.
We keep the E3 amplitude $a^{J=2}_3$ vanishing, which is the
consequence of the single quark radiation
hypothesis~\cite{Karl:1980wm,Olsson:1984zm}.

As was done in Ref.\cite{Ma:2010vd}, we have tried to improve the uncertainties in the ratio $r\equiv m_c^2\langle\mathcal{O}^{\chicz}(\so)\rangle/\langle\mathcal{O}^{\chicz}(\p01)\rangle$ by
using the LHCb data~\cite{LHCb:2012ac} and CMS data~\cite{Chatrchyan:2012ub}. With the Tevatron data, it was determined to be $r=0.27\pm0.06$. But its accuracy is not improved significantly with the updated LHC data. With the LHCb data~\cite{LHCb:2012ac}, $r$ varies from $0.35$ to $0.31$ when using a different $p_T$ cutoff. (To be compatible with our $\jpsi$ case~\cite{Chao:2012iv}, we always ignore the data when $p_T<7\rm{GeV}$.) Using the CMS data~\cite{Chatrchyan:2012ub}, we find $r$ has very weak dependence on $p_T$ cutoff, and its value is almost $0.25$ with unpolarized hypothesis. The substantial uncertainty in the $r$ extraction is due to different polarisation hypotheses. The $r$ value changes from $0.21$ to $0.31$ in two extreme hypotheses~\cite{Chatrchyan:2012ub}. Therefore, it is acceptable for us to choose $r=0.27\pm0.06$ here. Here, we may choose $r=0.27\pm0.06$ as an acceptable value, and the value of $r$ from different extractions are well embodied in its uncertainties. We emphasize further that measurements with higher resolution, especially in the high $p_T$ region, will be very useful to
improve our NRQCD predictions.

\begin{figure*}[!hbtp]
\begin{center}
\includegraphics*[scale=0.45]{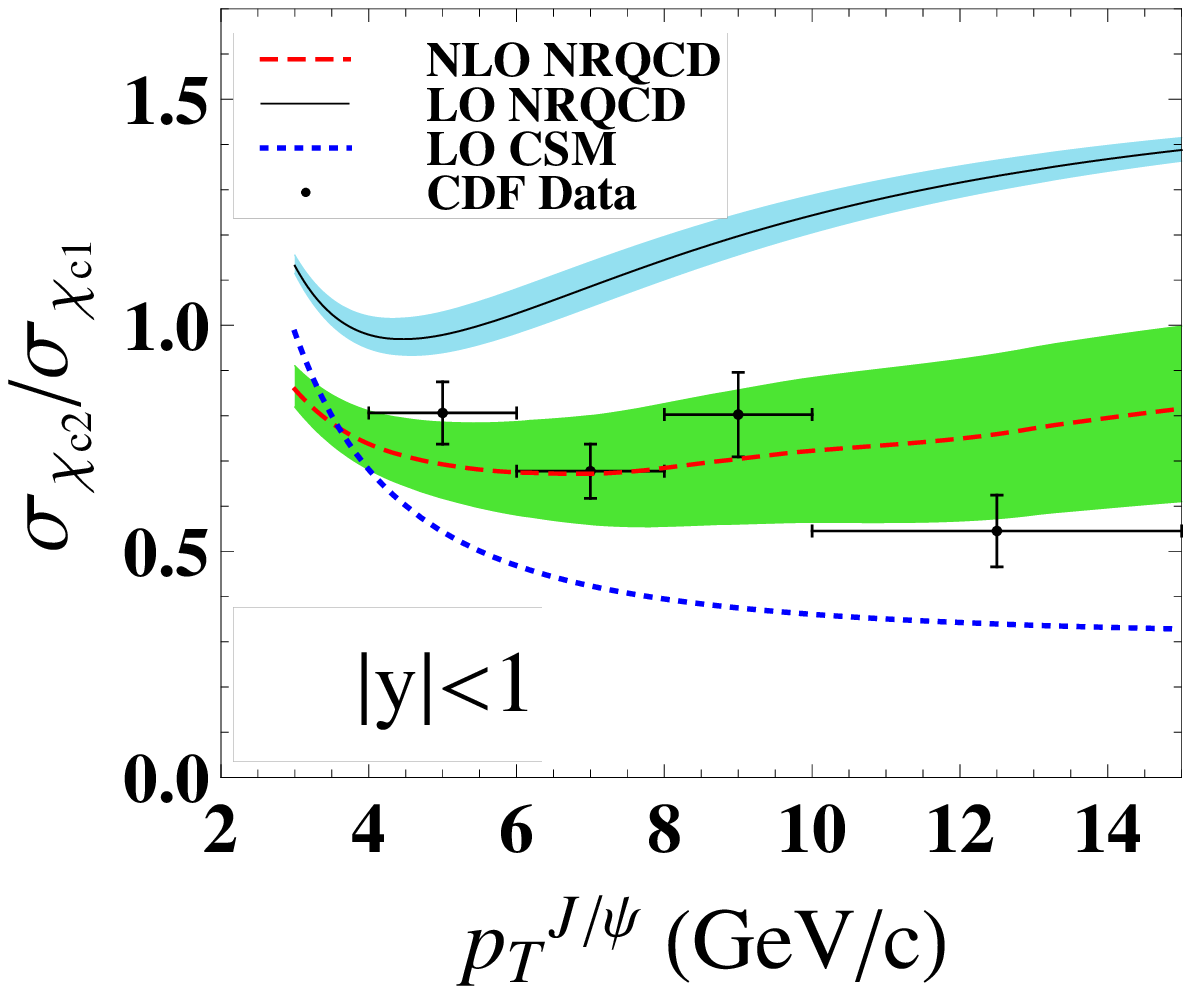}
\includegraphics*[scale=0.45]{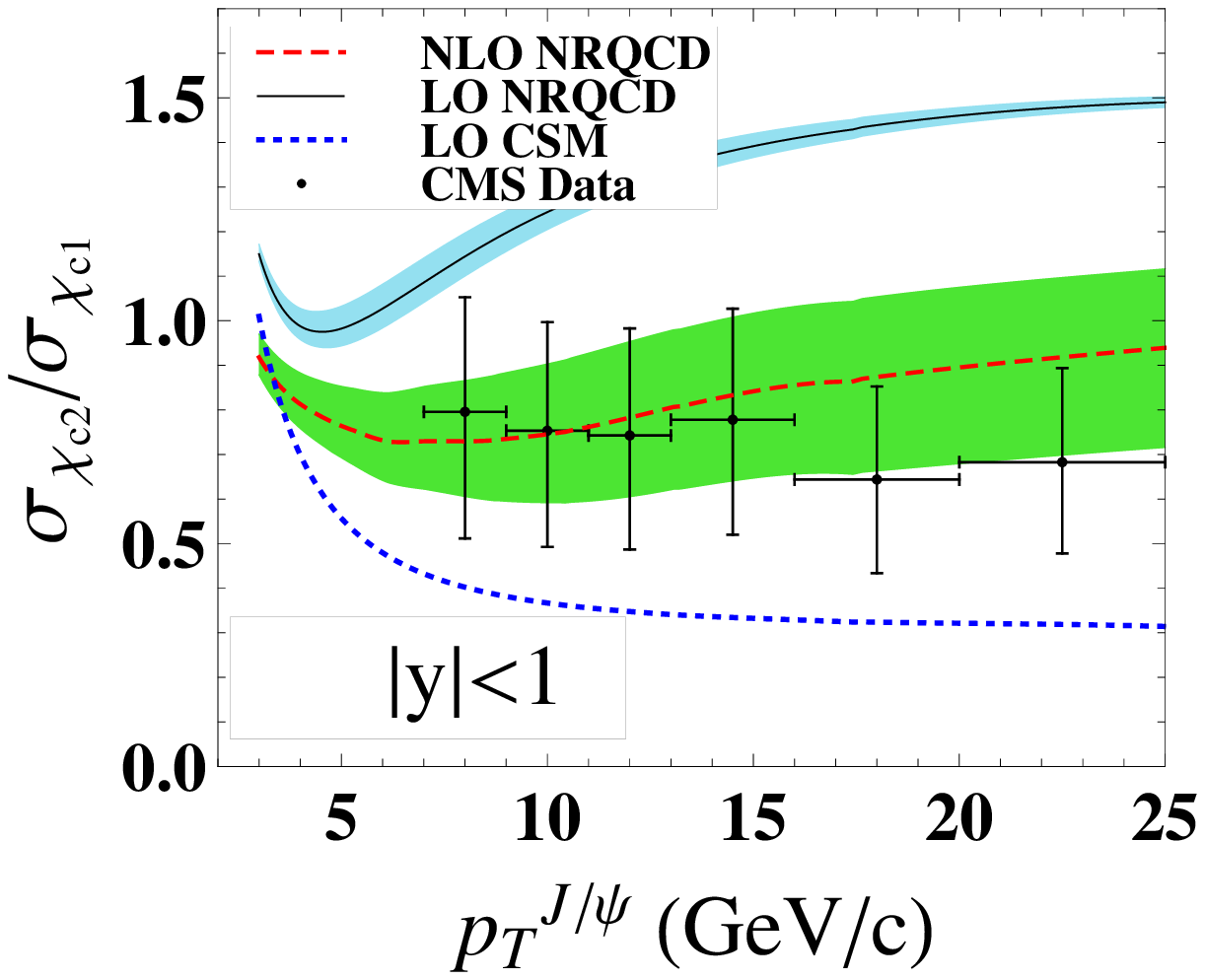}
\includegraphics*[scale=0.45]{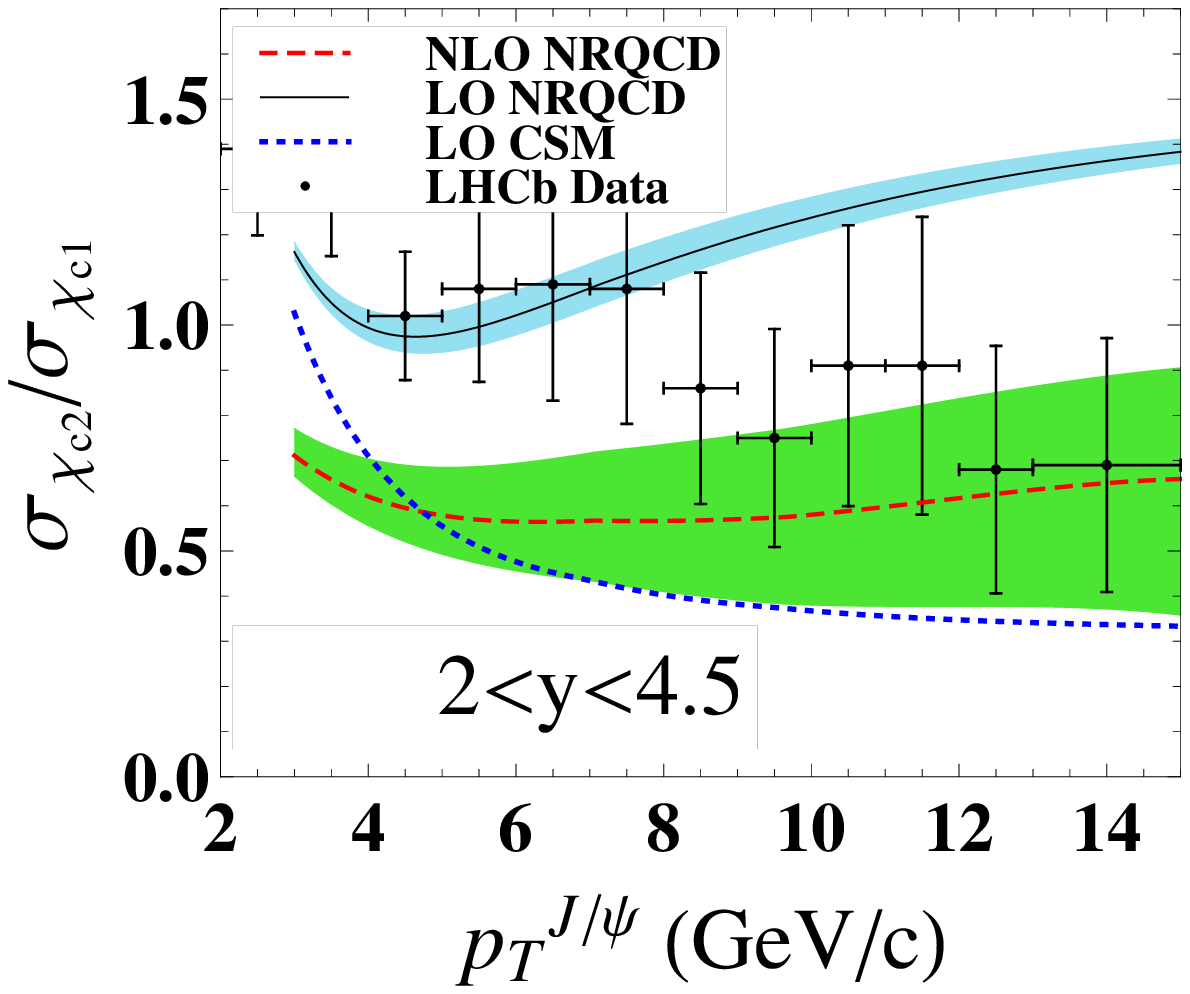}
\caption{\label{fig:ratio} (color online) The
cross-section ratio $\sigma_{\chict}/\sigma_{\chico}$ vs
the transverse momentum $p^{\jpsi}_T$  at the Tevatron Run
II (left panel) and LHC at $\sqrt{S}=7\rm{TeV}$ (right
two panels). The
rapidity cuts are the same as in the
experiments~\cite{Abulencia:2007bra,Chatrchyan:2012ub,LHCb:2012ac}. Results for
LO NRQCD (solid line), NLO NRQCD (dashed line) and LO CSM (dotted line) are shown.}
\end{center}
\end{figure*}

\begin{figure*}[!hbtp]
\begin{center}
\includegraphics*[scale=0.45]{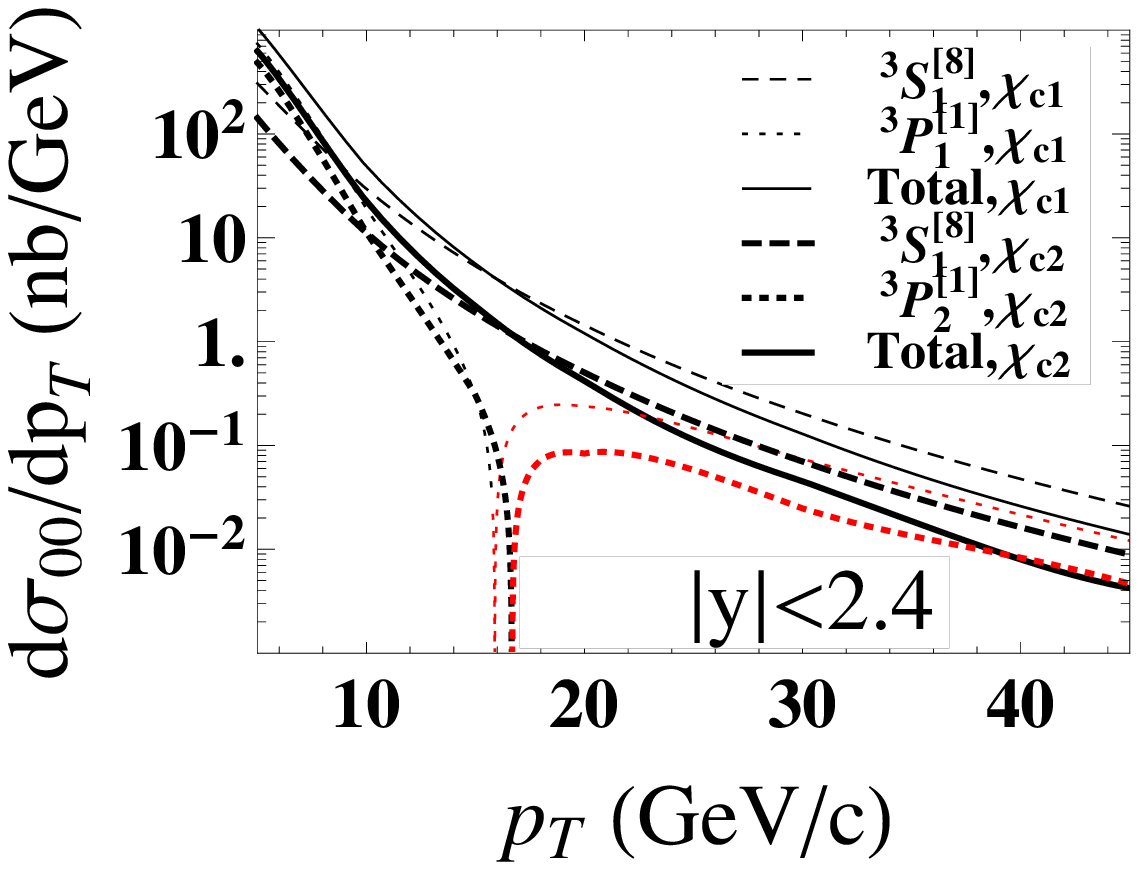}
\includegraphics*[scale=0.45]{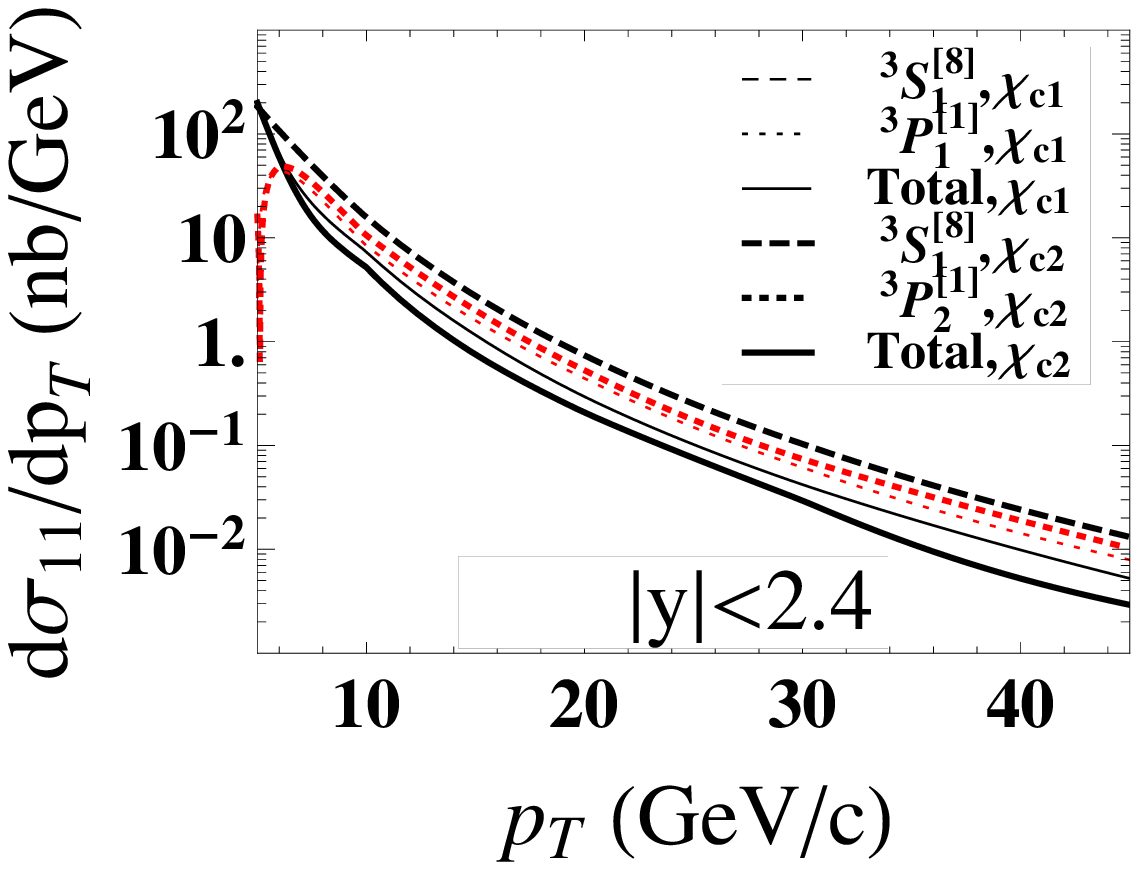}
\includegraphics*[scale=0.45]{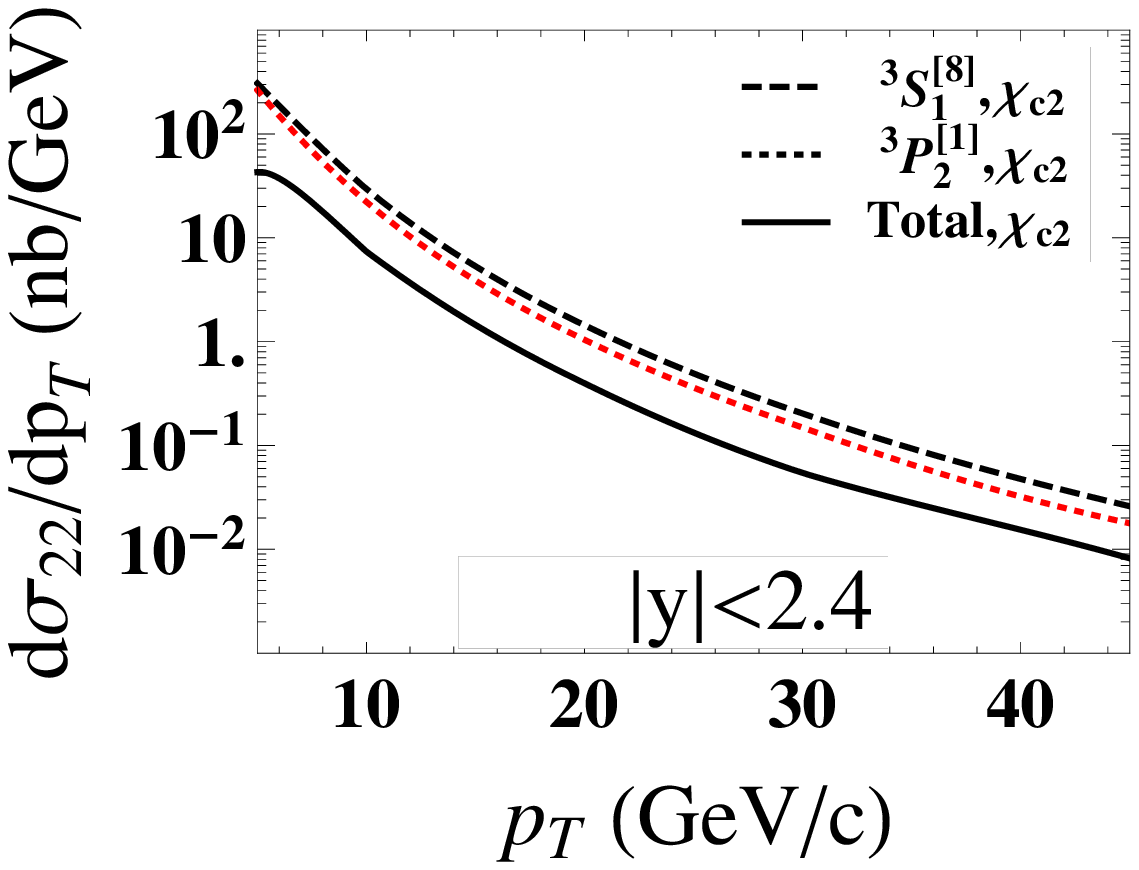}
\caption{\label{fig:dsigmapol} (color online) $d\sigma_{00}/dp_T$,$d\sigma_{11}/dp_T$, $d\sigma_{22}/dp_T$
for $pp\to\chicj+X(J=1,2)$ with $\sqrt{S}=7~\rm{TeV}$ and $|y|<2.4$ in the helicity frame at NLO in NRQCD.
The thin lines represent for $\chico$ whereas the thick lines represent for $\chict$. Negative values are marked red (lighter).}
\end{center}
\end{figure*}

\begin{figure}
\includegraphics[width=4.25cm]{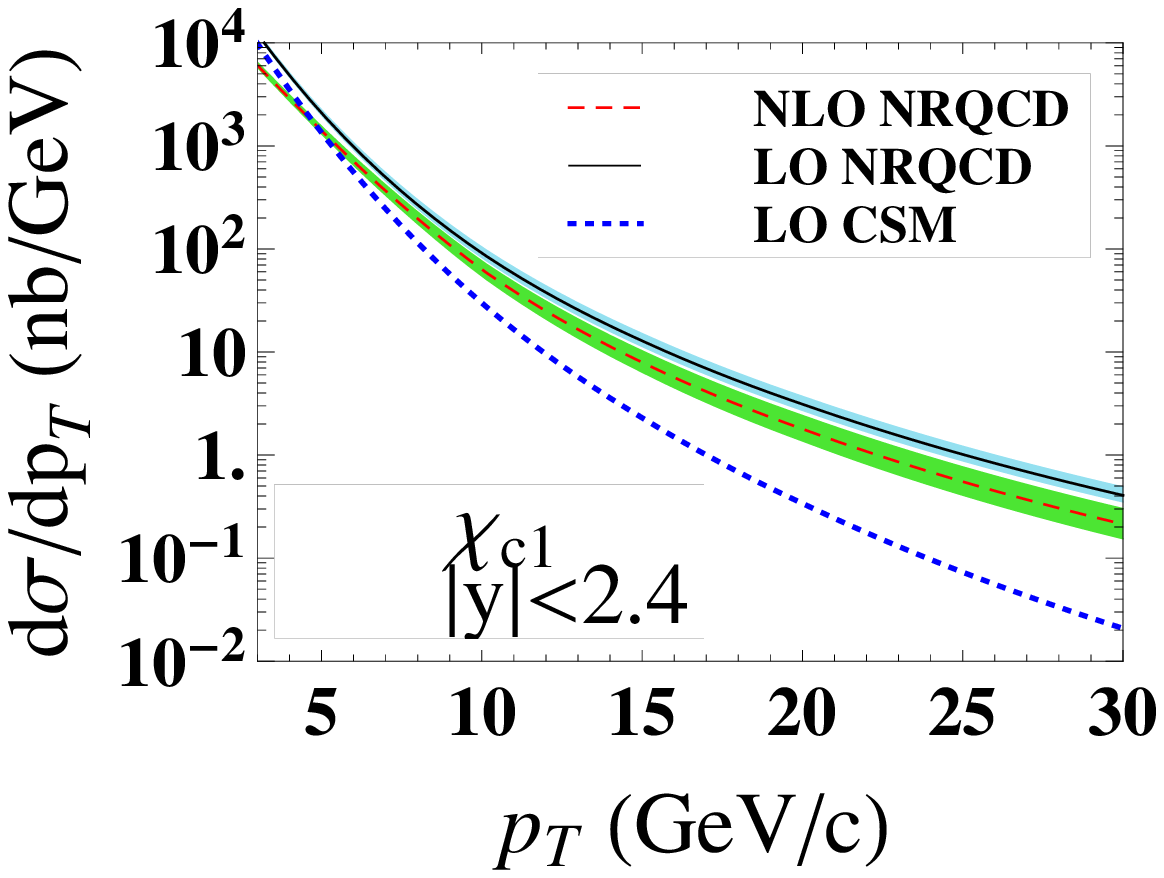}
\includegraphics[width=4.25cm]{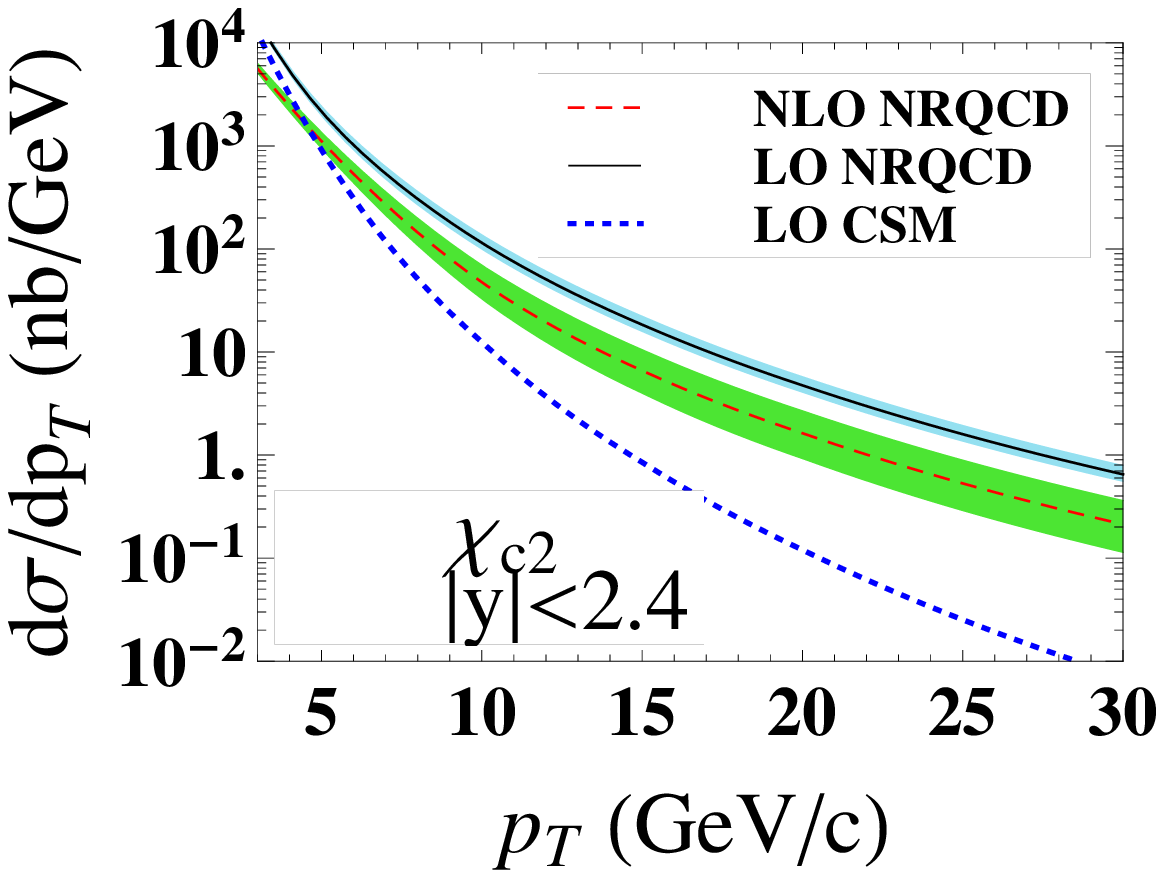}
\includegraphics[width=4.25cm]{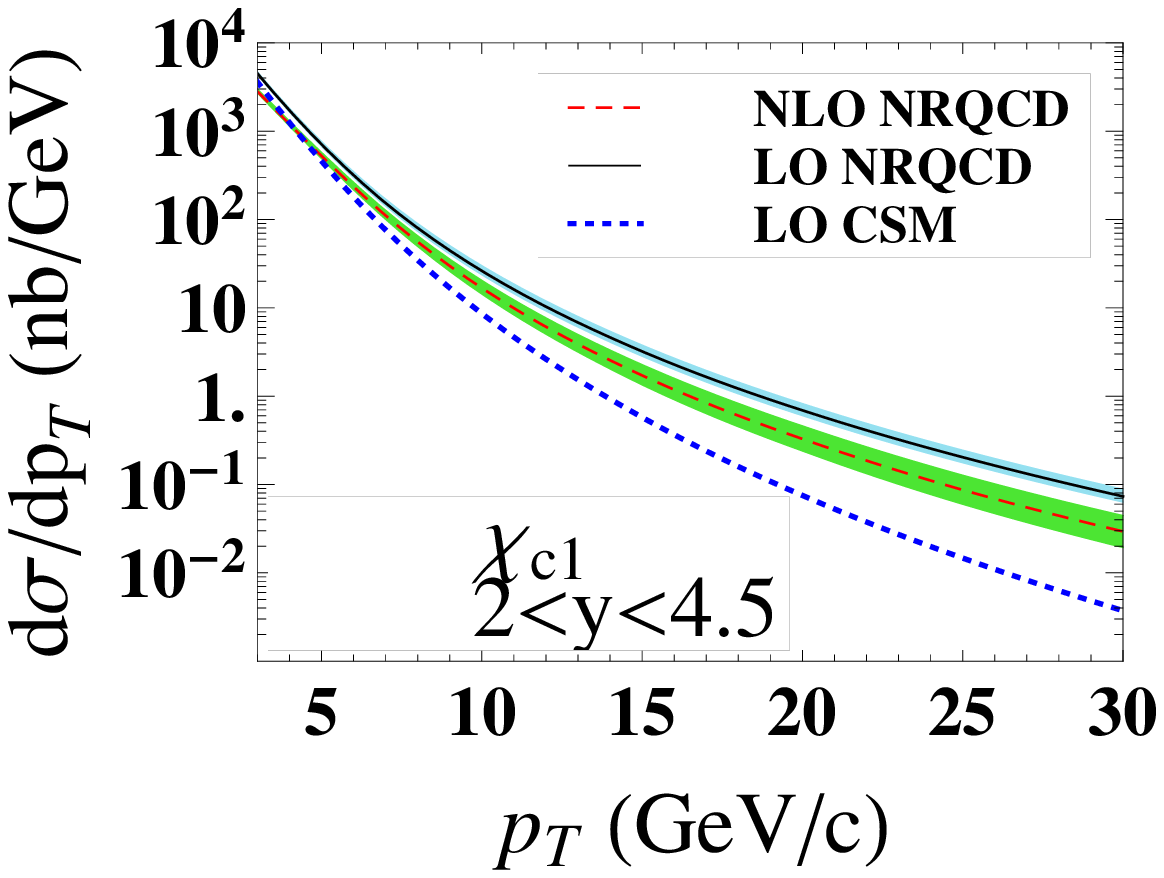}
\includegraphics[width=4.25cm]{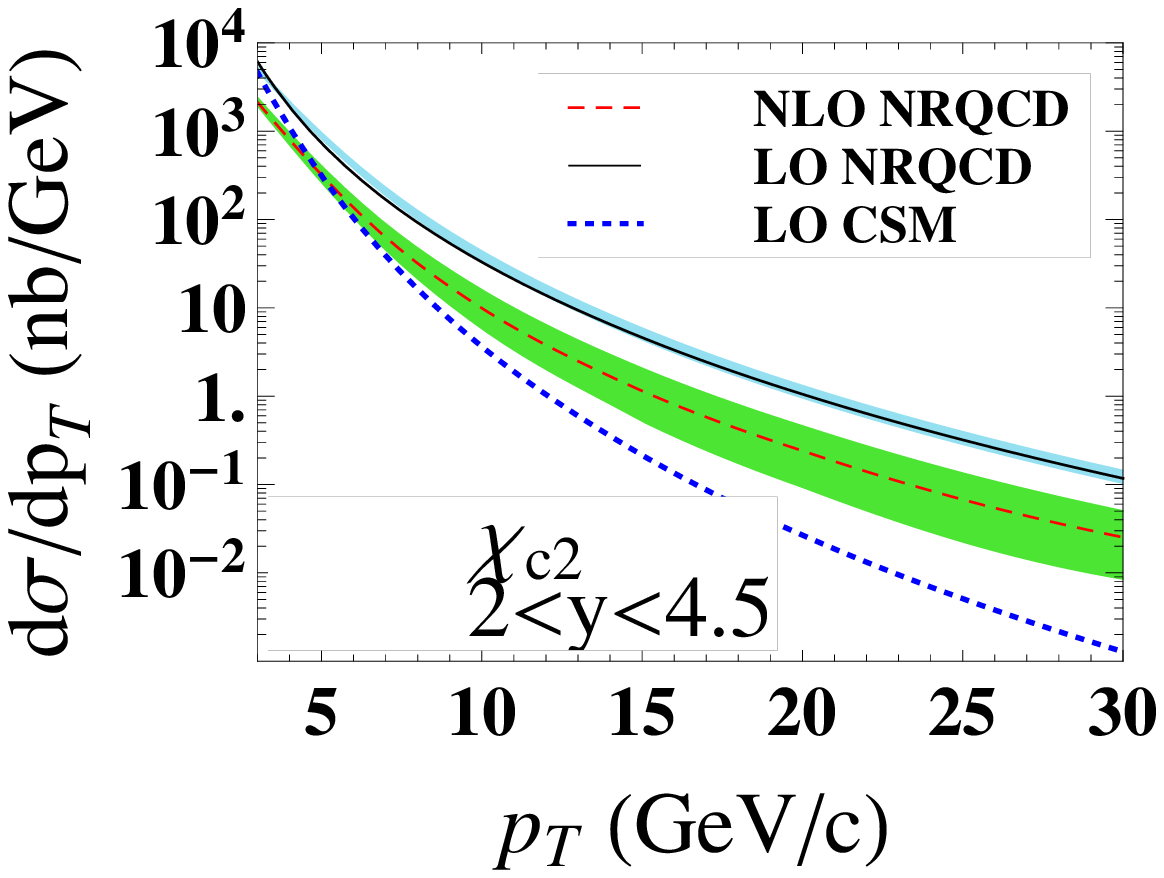}
\caption{\label{fig:cross} (color online) Predictions of
$p_T$ spectra for the helicity-summed $\chico$ (left column) and $\chict$
(right column) at the LHC with
$\sqrt{S}=7\rm{TeV}$.  Cross sections in the
central rapidity region ($|y|<2.4$) and forward rapidity region
($2<y<4.5$) for $\chi_c$ are plotted.  Results for LO NRQCD (solid
line), NLO NRQCD (dashed line) and LO CSM
(dotted line) are shown.}
\end{figure}

In Fig.\ref{fig:ratio}, the cross section ratios
$\sigma_{\chict}/\sigma_{\chico}$ at the Tevatron Run II and LHC are
shown. For comparison, besides the NLO NRQCD predictions, we also
plot the LO NRQCD results and the LO CSM results. We see
the NLO NRQCD results are consistent with the CDF
data~\cite{Abulencia:2007bra} and the CMS data~\cite{Chatrchyan:2012ub} in the whole $p^{\jpsi}_T$
region, while in the forward rapidity region the NLO
NRQCD prediction is in agreement with the LHCb
data~\cite{LHCb:2012ac} only when $p^{\jpsi}_T>8\rm{GeV}$, which may
imply that some unknown nonperturbative effects make our
fixed-order results unreliable when $p^{\jpsi}_T$ is lower.
Note that $p^{\jpsi}_T$ is obtained from
$p_T$ of $\chi_c$ by the mass
rescaling $p^{\jpsi}_T=\frac{m_{\jpsi}}{m_{\chicj}}p_T$, which
is proven to be a good approximation by the Monte Carlo simulation. Here the
masses $m_{\jpsi}=3.10{\rm GeV}, m_{\chico}=3.51{\rm GeV},
m_{\chict}=3.56{\rm GeV}$, and branching ratios
$\rm{Br}(\chico\to\jpsi\gamma)=0.344, \rm{Br}(\chict\to\jpsi\gamma)=0.195$
are taken from Ref.\cite{Nakamura:2010zzi}. We see also  that
the LO CSM prediction is substantially lower than the experimental
data. Two other important obstacles for CSM are  the measured cross
section of $\chicj$ at the Tevatron Run I~\cite{Ma:2010vd} and
ratio $\sigma(\chicj\to\jpsi\gamma)/\sigma(\jpsi)$ at
the LHC~\cite{LHCb:2012af}. While there are discrepancies between the LO CSM
predictions and the experimental data,  the
NLO NRQCD results are reasonably good. To present the predictions of
the cross sections at the LHC, we also show the corresponding curves
in Fig.\ref{fig:cross}. In Fig.\ref{fig:dsigmapol},we present the curves
of spin density matrix elements $d\sigma_{00}/dp_T,d\sigma_{11}/dp_T$(and $d\sigma_{22}/dp_T$)
for $\chico$($\chict$) with $\sqrt{S}=7$ TeV and $|y|<2.4$. To be more specific, we also show curves in different
Fock states( with LDMEs given above ). The negative values (see also Refs.\cite{Butenschoen:2012px,Chao:2012iv,Gong:2012ug}) are marked red.
\begin{table}[h]
\caption{\label{tab:limit} Upper and lower bound values of the
observables $\lambda_{\th}$ and $\lambda_{\th^{\prime}}$ for
$\chico$ and $\chict$.}
\begin{tabular}{c*{4}{c}}
\hline\hline \itshape ~\rm{Observable}~ & ~$\lambda^{\chico}_{\th}$~
& ~$\lambda^{\chict}_{\th}$~ & ~$\lambda^{\chico}_{\th^{\prime}}$~&
~$\lambda^{\chict}_{\th^{\prime}}$~
\\\hline ~\rm{Upper bound}~&~$0.556$~& $1.61$ & $0.994$ & $0.928$
\\ ~\rm{Lower bound}~&~$-0.217$~& $-0.803$ & $-0.332$ & $-0.574$
\\\hline\hline
\end{tabular}
\end{table}
\begin{figure}
\includegraphics[width=4.25cm]{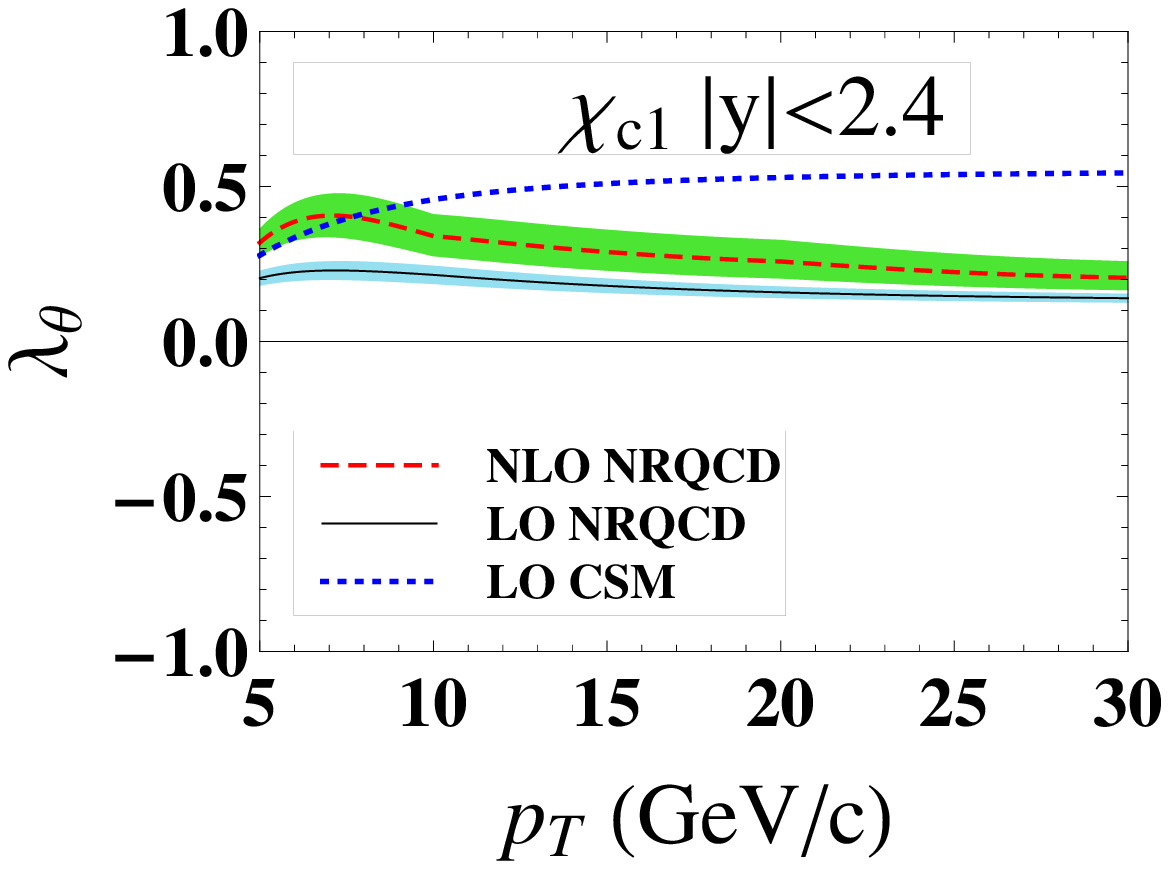}
\includegraphics[width=4.25cm]{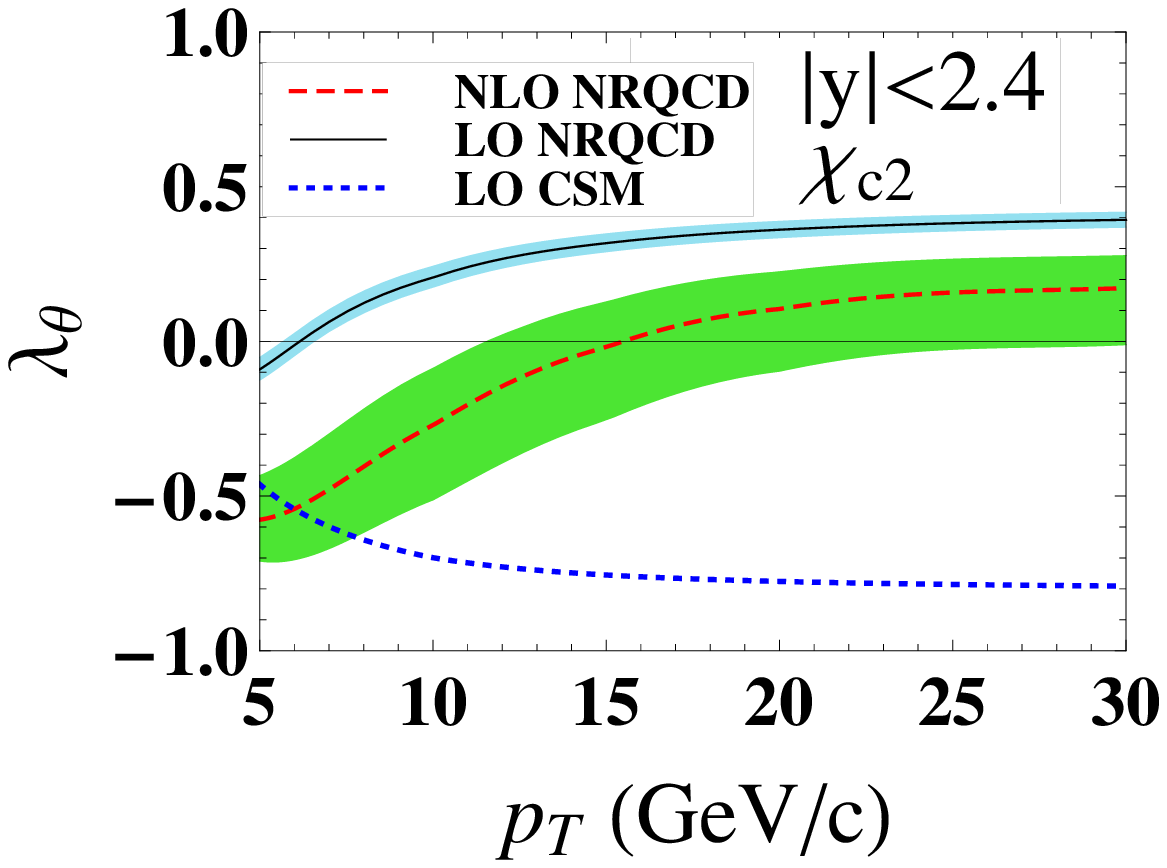}
\includegraphics[width=4.25cm]{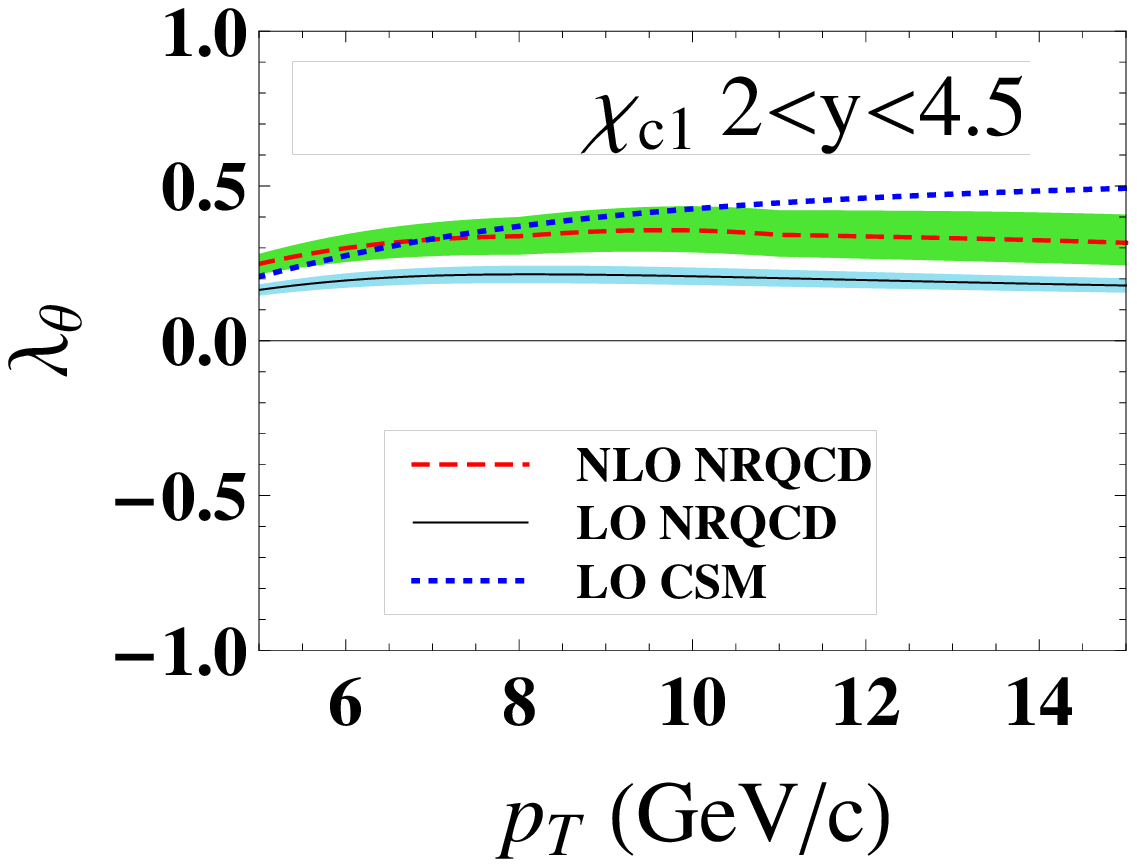}
\includegraphics[width=4.25cm]{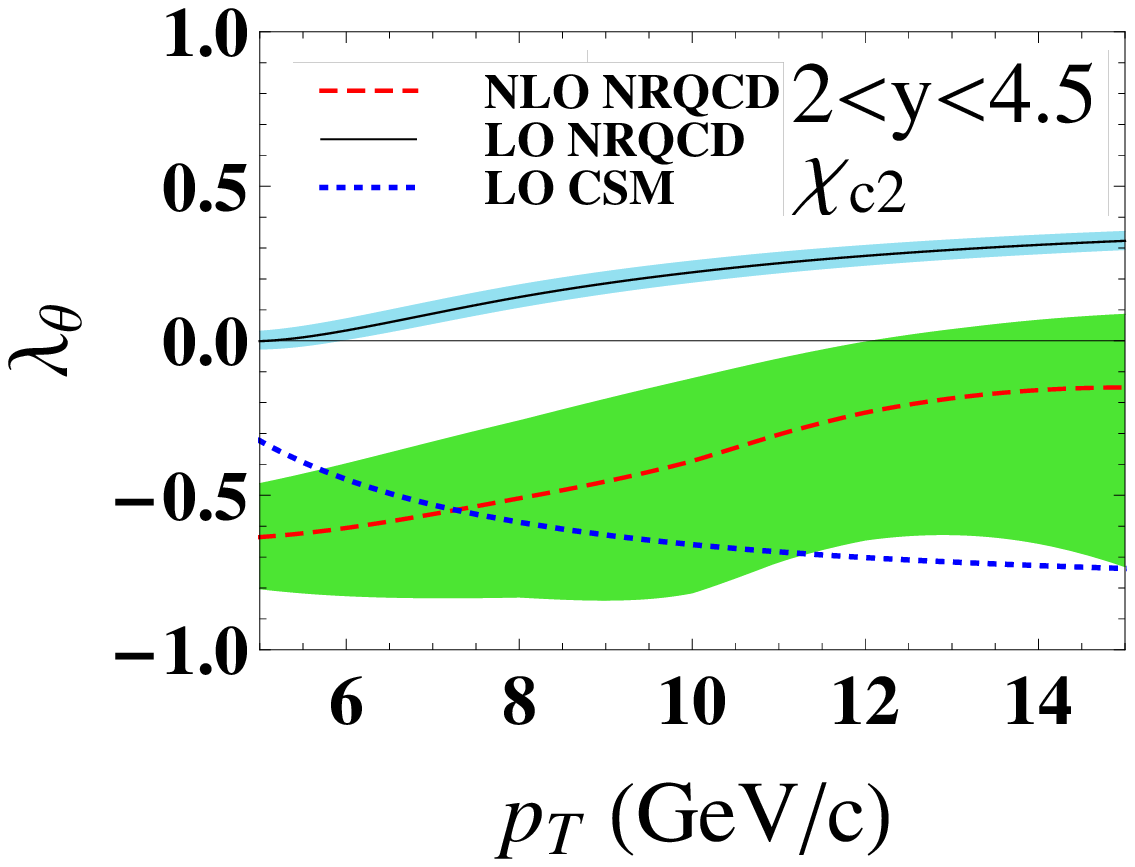}
\caption{\label{fig:thpol} (color online) The $p_{T}$ dependence of
$\lambda_{\th}$ with $\jpsi$ angular distributions from
radiative decays $\chico\to\jpsi\gamma$ (left column) and
$\chict\to\jpsi\gamma$ (right column) in the helicity
frame at the LHC with $\sqrt{S}=7\rm{TeV}$.  Results in central and forward rapidity
regions are plotted. The LO NRQCD (solid line),
NLO NRQCD (dashed line), and LO CSM (dotted line) predictions are
shown.}
\end{figure}
\begin{figure}
\includegraphics[width=4.25cm]{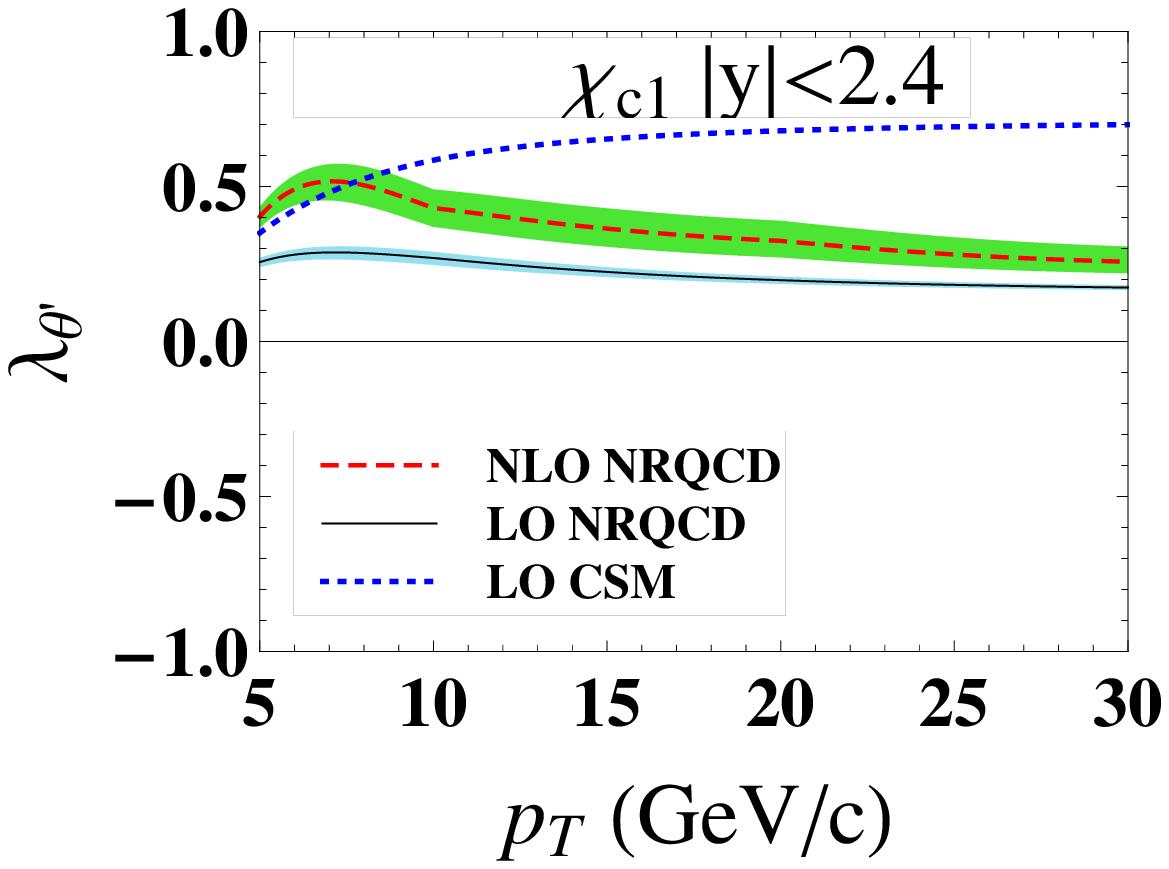}
\includegraphics[width=4.25cm]{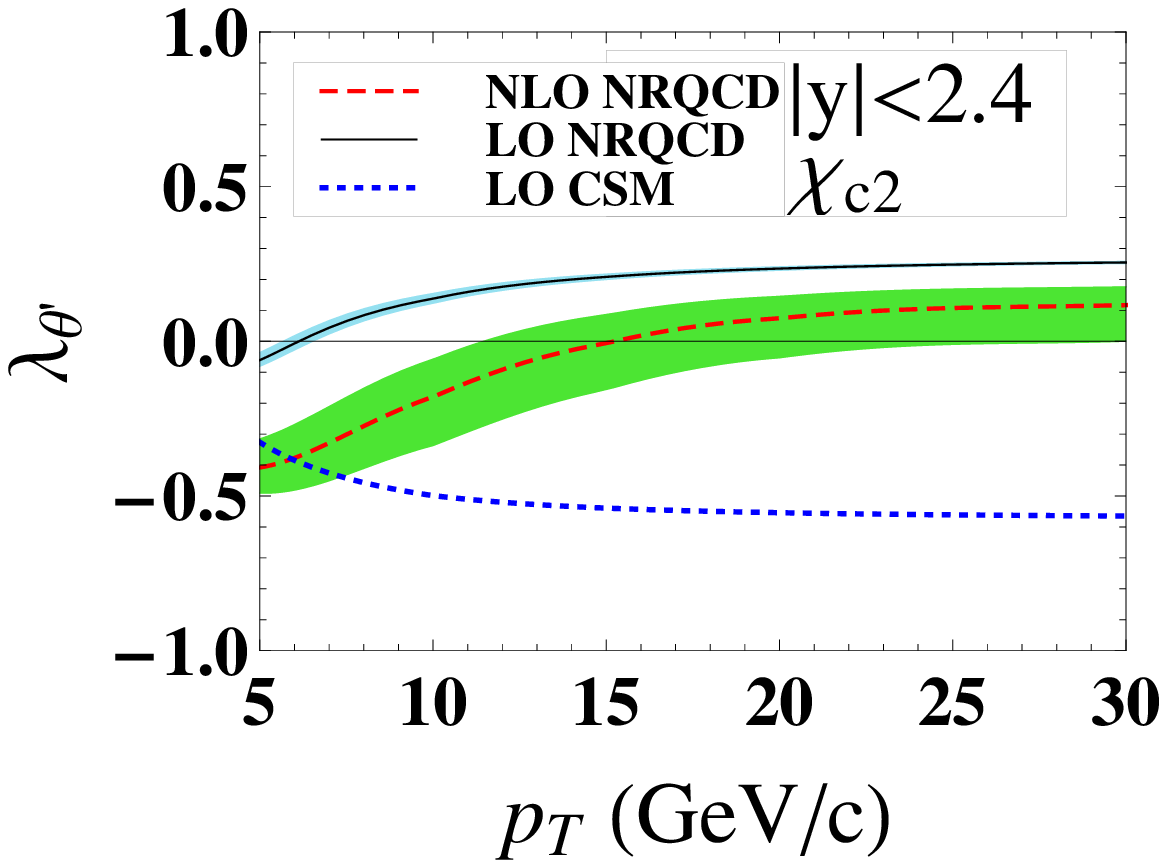}
\includegraphics[width=4.25cm]{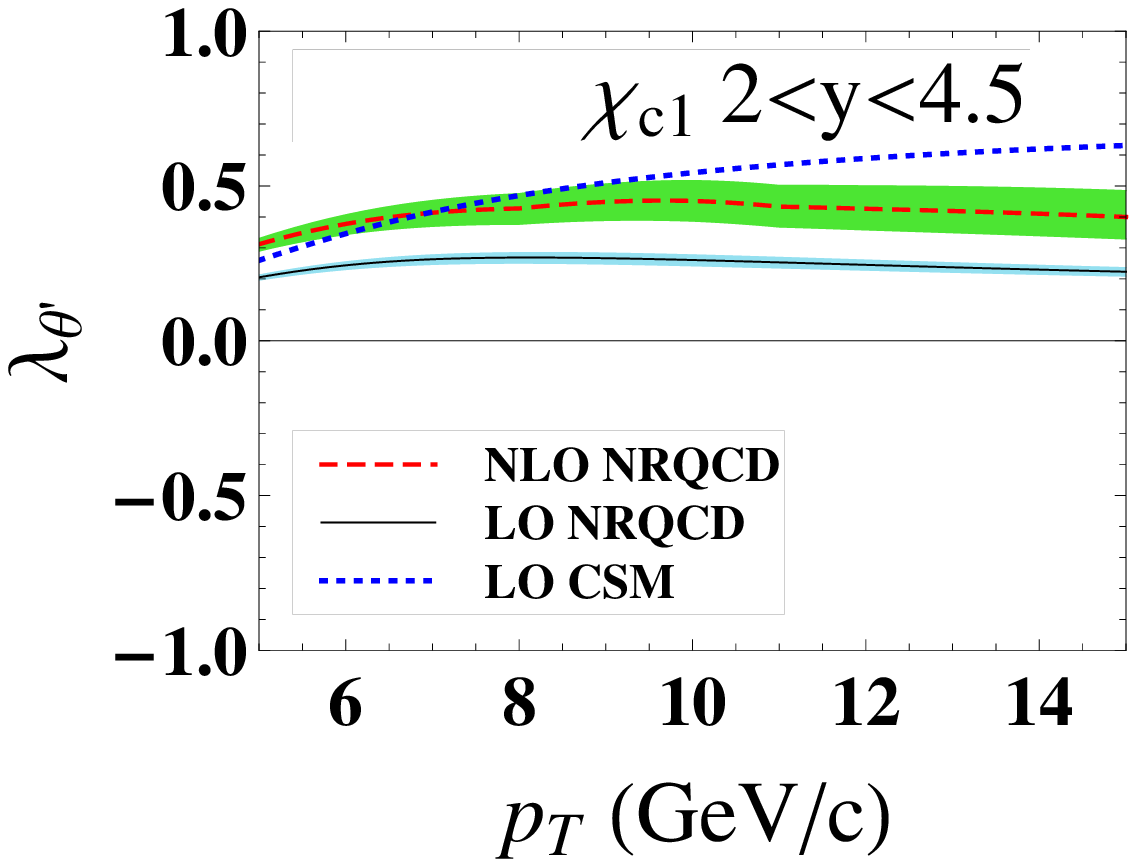}
\includegraphics[width=4.25cm]{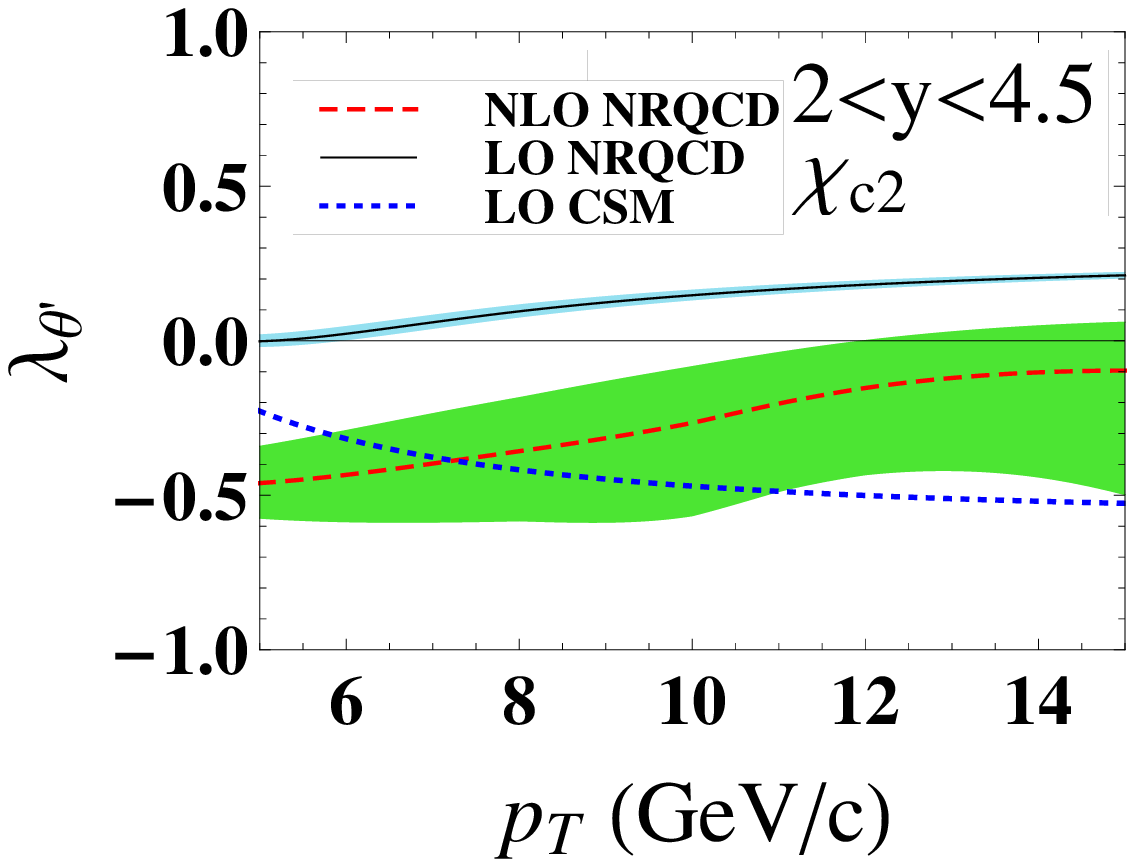}
\caption{\label{fig:thppol} (color online) The $p_{T}$ dependence of
$\lambda_{\th^{\prime}}$ with  dilepton angular
distributions from cascade decays $\chico\to\jpsi\gamma\to
l^+l^-\gamma$ (left column) and $\chict\to\jpsi\gamma\to
l^+l^-\gamma$ (right column) in the helicity frame at
the LHC with $\sqrt{S}=7\rm{TeV}$. Results in central rapidity
and forward rapidity regions are plotted, and the LO NRQCD
(solid line), NLO NRQCD (dashed line) and LO CSM (dotted line)
predictions are shown.}
\end{figure}

For the numerical results of the
polarization observables of $\chico$ and $\chict$, we use
expressions in Eqs.(\ref{eq:th1},\ref{eq:th2},and \ref{eq:thp}) and obtain , first, the lower and upper bound values of $\lambda_{\th}$
and $\lambda_{\th^{\prime}}$ for $\chi_c$ regardless of its
production mechanisms. They are presented in Table \ref{tab:limit}.
When
$\rho^{\chico}_{1,1}=\rho^{\chico}_{-1,-1}\ll\rho^{\chico}_{0,0}$, the
polar observables for $\chico$ approach their maximal values,
whereas the minimal values are obtained when
$\rho^{\chico}_{1,1}=\rho^{\chico}_{-1,-1}\gg\rho^{\chico}_{0,0}$.
For $\chict$,  the polar asymmetry coefficients $\lambda_{\th}$ and
$\lambda_{\th^{\prime}}$ are maximum when
$\rho^{\chict}_{2,2}=\rho^{\chico}_{-2,-2}\gg\rho^{\chict}_{1,1}=\rho^{\chict}_{-1,-1},\rho^{\chict}_{0,0}$
and minimum when
$\rho^{\chict}_{2,2}=\rho^{\chico}_{-2,-2},\rho^{\chict}_{1,1}=\rho^{\chict}_{-1,-1}\ll\rho^{\chict}_{0,0}$.
The $p_T$ distributions of $\lambda_{\th}$ and
$\lambda_{\th^{\prime}}$ are shown in Figs. \ref{fig:thpol} and
\ref{fig:thppol}, respectively. It is worth noting that the
transformation relation between the spin density matrices of $\so$
and those of $\pjs$~\cite{Shao:2012fs}
\begin{align}
{\rho}^{\so\to\chicj}_{J_z,J_z^{\prime}}& \propto
\sum_{l_z,s_z,s_z^{\prime}=\pm1,0}\rho^{\so}_{s_z,s_z^{\prime}} \nonumber\\ \times & \langle1,l_z;1,s_z|J,J_z\rangle  \langle1,l_z;1,s_z^{\prime}|J,J_z^{\prime}\rangle\,,
\end{align}
is used in our numerical results. The error bands in these figures are due to
uncertainties of the CO LDMEs
$\langle\mathcal{O}^{\chicj}(\so)\rangle$ and errors in the
normalized multipole amplitudes. From
Figs.\ref{fig:thpol} and \ref{fig:thppol}, we see that the
measurements of these polarization observables may provide another
important way to test the CO mechanism in the hadroproduction of
heavy quarkonium.

In summary, we have performed an analysis of the polarized $\chico$
and $\chict$ production at the LHC in NRQCD and in the
color-singlet model. The complete NLO NRQCD predictions are given for the first
time. These observables may provide important
information, which is not available in the helicity-summed $p_T$ spectra,
in testing the validity of NRQCD factorization.
Compared with $J/\psi$ production, the prompt $\chi_c$ production may play a unique role in understanding the heavy quarkonium production mechanism. Therefore, we propose to measure these polarization observables at the LHC.

We are grateful to C.~Meng, Y.J.~Zhang and H.~Han for helpful discussions.
This work was supported in part by the National
Natural Science Foundation of China
(Nos.11021092,11075002). Y.-Q.M is supported by the U.S.
Department of Energy, contract number DE-AC02-98CH10886.




\providecommand{\href}[2]{#2}\begingroup\raggedright
\endgroup

\end{document}